# Fermi-surface reconstruction by stripe order in cuprate superconductors

F. Laliberté[1], J. Chang[1], N. Doiron-Leyraud[1], E. Hassinger[1], R. Daou[1,†], M. Rondeau[1], B. J. Ramshaw[2], R. Liang[2,3], D. A. Bonn[2,3], W. N. Hardy[2,3], S. Pyon[4], T. Takayama[4], H. Takagi[4,5], I. Sheikin[6], L. Malone[7], C. Proust[3,7], K. Behnia[8] & Louis Taillefer[1,3]

*1 Département de physique and RQMP, Université de Sherbrooke, Sherbrooke, Québec J1K 2R1, Canada*

*2 Department of Physics and Astronomy, University of British Columbia, Vancouver, British Columbia V6T 1Z4, Canada*

*3 Canadian Institute for Advanced Research, Toronto, Ontario M5G 1Z8, Canada*

*4 Department of Advanced Materials, University of Tokyo, Kashiwa 277-8561, Japan*

*5 RIKEN (The Institute of Physical and Chemical Research), Wako, 351-0198, Japan*

*6 Laboratoire National des Champs Magnétiques Intenses, Grenoble 38042, France*

*7 Laboratoire National des Champs Magnétiques Intenses, UPR 3228, (CNRS-INSA-UJF-UPS), Toulouse 31400, France*

*8 LPEM (UPMC-CNRS), ESPCI, Paris 75231, France*

**The origin of pairing in a superconductor resides in the underlying normal state. In the cuprate high-temperature superconductor YBa$_2$Cu$_3$O$_y$ (YBCO), application of a magnetic field to suppress superconductivity reveals a ground state that appears to break the translational symmetry of the lattice, pointing to some density-wave order. Here we use a comparative study of thermoelectric transport in the cuprates YBCO and La$_{1.8-x}$Eu$_{0.2}$Sr$_x$CuO$_4$ (Eu-LSCO) to show that the two materials exhibit the same process of Fermi-surface reconstruction as a function of temperature and doping. The fact that in Eu-LSCO this reconstruction coexists with spin and charge modulations that break translational symmetry shows that stripe order is the generic non-superconducting ground state of hole-doped cuprates.**

† Present address : Max Planck Institute for Chemical Physics of Solids, 01187 Dresden, Germany

In the underdoped region of the phase diagram, quantum oscillations have revealed a small Fermi surface pocket in the cuprate superconductor YBCO (ref. 1), in contrast to the large Fermi surface observed on the overdoped side. This implies that there is a quantum critical point (QCP) near optimal doping where the Fermi surface of the non-superconducting ground state is reconstructed. Understanding this QCP is essential, and if it corresponds to the onset of some density-wave order that breaks the translational symmetry of the $CuO_2$ planes, then the fluctuations of that order may well cause both the linear temperature dependence of the "strange-metal phase" and the pairing[2]. Such a density-wave scenario applies to organic superconductors[3], for example, and possibly also to iron-pnictide and heavy-fermion superconductors[4]. In cuprates, the nature and origin of the small Fermi pocket are the subject of much debate[5]. An electron-like rather than hole-like pocket would strongly favour a density-wave scenario whereby the Fermi surface undergoes a reconstruction driven by the new periodicity[6], as in the case of antiferromagnetism[7], $d$-density-wave order[8], or stripe order[9].

Here we report quantum oscillations in the Seebeck and Nernst coefficients of YBCO and show, from the magnitude and sign of the Seebeck coefficient, that they come from an electron pocket. Using measurements of the Seebeck coefficient as a function of hole doping $p$, we show that the evolution of the Fermi surface in YBCO is the same as in Eu-LSCO, a cuprate where stripe order – a modulation of spin and charge densities[10,11] – is well established[12,13,14,15]. The electron pocket is most prominent where stripe order is strongest, at $p = 1/8$. This shows that Fermi-surface reconstruction is a generic mechanism of underdoped cuprates, intimately related to stripe order.

## Results

**Quantum oscillations.** Quantum oscillations in YBCO at a hole doping $p \approx 0.1$ have a dominant frequency $F \approx 530$ T (refs. 1, 16, 17), giving a closed Fermi surface area 30 times smaller than that found in overdoped $Tl_2Ba_2CuO_{6+y}$ at $p \approx 0.3$ (ref. 18). A key question is the sign of the associated carriers. The large negative Hall coefficient $R_H$ observed in the same YBCO samples at low temperature is evidence of electron-like carriers[19,20]. However, because a negative Hall resistance can in principle come from moving vortices or negative curvature in a hole-like Fermi surface, it is important to confirm the sign of carriers using evidence insensitive to these effects. This can be done by measuring the Seebeck coefficient (or thermopower) $S$, a standard measure of carrier sign. In Figure 1a, we show $S$ and the Nernst coefficient $v$ of YBCO at $p = 0.11$, measured at low temperature as a function of magnetic field $B$ up to 28 T. The fact that $v$ is flat at high field shows that vortex contributions have become negligible above 25 T, and the normal state has for all practical purposes been reached. The fact that $S$ is negative in that high-field normal state confirms that the dominant carriers are electron-like.

A zoom on the data at 2 K (Figure 1b) reveals quantum oscillations in both $S$ and $v$, proving the presence of a small closed pocket in the Fermi surface of YBCO in fields as low as 24 T. The frequency of the oscillations (in $1/B$) is $F \approx 520$ T, in agreement with quantum oscillations in the resistance[1] and magnetization[16,17] of similar YBCO crystals. Given a cyclotron mass $m^* = 1.76 \pm 0.07\ m_0$ (ref. 16), we obtain the Fermi temperature $T_F = (e\ \hbar\ /\ k_B)\ (F\ /\ m^*) = 410 \pm 20$ K, where $m_0$ and $e$ are the electron mass and charge, respectively, and $k_B$ is the Boltzmann constant. From $T_F$ we estimate the magnitude of the Seebeck and Nernst coefficients expected for this Fermi pocket at $T \to 0$, using standard expressions known to agree well with experiment in several correlated electron systems. These yield (ref. 21) :

$$S\ /\ T \approx -(\pi^2\ /\ 3)\ (3/2 + \zeta)\ (k_B\ /\ e)\ (1\ /\ T_F) = -0.9 \pm 0.2\ \mu V\ /\ K^2 \quad, \qquad (1)$$

where the negative sign is for electron-like carriers. The upper and lower bounds on the uncertainty correspond to assuming an energy-independent relaxation time ($\zeta = 0$) or mean-free path ($\zeta = -1/2$), respectively[21]. The measured value at low temperature is $S\ /\ T = -0.8 \pm 0.1\ \mu V\ /\ K^2$ (Figure 1a). The magnitude of the quasiparticle Nernst coefficient at $T \to 0$, $v\ /\ T = -13 \pm 3$ nV / $K^2$ T (Fig. 1a), is also the expected value[22]

$$|\ v\ /\ T\ | \approx (\pi^2\ /\ 3)\ (k_B\ /\ e)\ (\mu\ /\ T_F) = 13\ \pm 3\ nV\ /\ K^2\ T \quad, \qquad (2)$$

given the mobility $\mu = 0.02 \pm 0.006$ $T^{-1}$ obtained from quantum oscillations[16]. This excellent quantitative agreement shows that in YBCO at $p = 0.11$ the quantum oscillations come from a small electron-like Fermi pocket of high mobility.

**Seebeck coefficient.** In Figure 2a and 2b, the normal-state Seebeck coefficient of YBCO and Eu-LSCO, respectively, is plotted as a function of temperature for different dopings. We first discuss the YBCO data. At all dopings, $S\ /\ T$ is positive at high temperature. For $p = 0.09$, 0.10, 0.11 and 0.12, $S\ /\ T$ starts to drop monotonically below $\sim 80$ K, to become negative in the $T = 0$ limit, changing sign at a temperature $T_0^S$. We infer that an electron pocket is present in the Fermi surface at all four dopings. By contrast, at $p = 0.08$, $S\ /\ T$ shows no downturn down to the lowest temperature. The same evolution was reported recently in the Hall coefficient, with $R_H(T)$ crossing to negative values at a sign-change temperature $T_0^H$ when $p > 0.08$ and remaining positive at $p = 0.08$ (ref. 20). The sudden qualitative change in $R_H(T)$ across $p = 0.08$, now reproduced in $S(T)$, is attributed to a change in Fermi surface topology whereby the electron pocket disappears below $p = 0.08$ (ref. 20).

A convenient way to picture the doping evolution of the Fermi surface is to plot $T_0^S$ on a phase diagram, as shown in Figure 3a for YBCO, and Figure 3b for Eu-LSCO. For YBCO, $T_0^S$ and $T_0^H$ are seen to track each other closely, growing monotonically and in parallel from zero at $p = 0.08$ to their maximal value at $p = 0.12$. Hall measurements up to 60 T reveal that $T_0^H$ comes down at higher doping, so that $T_0^H$ forms a dome starting at $p = 0.08$, peaking at $p = 1/8$, and ending at $p \approx 0.17$ (ref. 20). This dome is the region of the phase diagram where the electron pocket dominates the transport properties.

Two questions arise: Is this Fermi-surface evolution unique to YBCO? What causes the Fermi surface to reconstruct below $\sim 80$ K ? We address these by turning to Eu-LSCO, a cuprate superconductor with a different crystal structure: tetragonal rather than orthorhombic, with single rather than double $CuO_2$ layers, without CuO chains, in which doping is done by substituting Sr rather than adding oxygen. In Figure 4, we compare the normal-state Seebeck coefficient of Eu-LSCO to that of YBCO at the same doping, $p = 0.11$. The two curves are essentially identical: $S / T$ has the same positive value at high temperature, it starts to drop at the same onset temperature near 80 K, it crosses zero at the same temperature $T_0^S = 40$ K, and it reaches a large negative value at $T \to 0$ in both cases. (Since $S$ is an intensive quantity, its magnitude is independent of the number of $CuO_2$ planes per unit cell.)

Our main finding is that the doping evolution of $S / T$ in Eu-LSCO is fundamentally identical to that of YBCO: at $p = 0.10$, 0.11 and 0.12, it drops to become negative, while at $p = 0.08$ it remains positive down to low temperature (Figure 2b). As shown in Figure 5 for Eu-LSCO, a drop to a negative $S / T$ is also observed at $p = 0.16$, while for $p = 0.21$ and 0.24 no drop is seen and $S / T$ is positive at all temperatures. In Figure 3, a plot of $T_0^S$ vs $p$ captures the striking similarity between Eu-LSCO and YBCO. We conclude that the Fermi surface of these two cuprates undergoes the same reconstruction as a function of temperature and doping, pointing to a universal phenomenon amongst hole-doped cuprates. This immediately implies that quantum oscillations and the downturns in $S / T$ and $R_H$ come from a Fermi pocket associated with the $CuO_2$ planes (and not the CuO chains of YBCO, for example).

## Discussion

Upon cooling below $\sim 100$ K, Eu-LSCO undergoes an ordering process called "stripe order"[10,11], also observed in the closely-related material $La_{1.6-x}Nd_{0.4}Sr_xCuO_4$ (Nd-LSCO), which involves both charge[15,23] and spin modulations[12,13,24,25]. These are in general incommensurate with the lattice, with a period of approximately $4a$ and $8a$ at $p \approx 1/8$, respectively, where $a$ is the in-plane lattice constant. The charge-ordering temperature $T_{CO}$ in Eu-LSCO is displayed in Figure 3b. At $T \to 0$, both modulations are present throughout the doping range $0.10 \leq p \leq 0.15$ (refs. 12, 13, 15), and by extrapolation from $p \approx 0.08$ to beyond $p \approx 0.2$. Because they break the translational symmetry of the lattice, these stripe modulations will inevitably cause a reconstruction of the Fermi surface. Calculations show that the reconstructed Fermi surface will in general contain electron pockets[9]. We have now shown that throughout the doping range $0.10 \leq p \leq 0.16$, the Fermi surface of Eu-LSCO does indeed contain an electron pocket. We conclude that stripe order is the mechanism responsible for Fermi-surface reconstruction in Eu-LSCO. By close analogy, we infer that the same mechanism acts in YBCO. By extension, Fermi-surface reconstruction by stripe order is likely to be a generic property of hole-doped cuprates in the approximate range $0.08 < p < 0.2$, and possibly beyond.

Apart from the presence of a small closed electron pocket, little is known with certainty about the reconstructed Fermi surface of YBCO or Eu-LSCO. In particular, the $k$-space location of the electron pocket and the existence of other Fermi-surface sheets are open questions. Studies of the electronic structure by angle-resolved photoemission spectroscopy in Eu-LSCO (ref. 26) and Nd-LSCO (ref. 27) at $p \approx 1/8$, performed between $T_c$ and $T_{CO}$, revealed evidence of a Fermi-surface reconstruction. However it is unclear whether these effects are due to stripe order or to structural distortions, and no direct evidence for an electron pocket was found.

The excellent agreement between the measured value of $S / T$ at $T \to 0$ and that calculated assuming the Fermi surface consists of only the electron pocket implies that other Fermi-surface sheets, as yet unobserved, must have a much lower conductivity. In particular, a hole-like sheet would make the Seebeck and Hall coefficients less negative. From the frequency of quantum oscillations in YBCO at $p \approx 0.1$, the Hall coefficient of the electron pocket alone should be $R_e = -1 / n\,e = -15 \text{ mm}^3 / \text{C}$, since $n = F / \Phi_0 = 0.038$ carriers per planar Cu atom[19], assuming an isotropic pocket. The measured value of the Hall coefficient is $R_H = -35 \text{ mm}^3 / \text{C}$ (ref. 19). The fact that $|R_H| > |R_e|$ suggests that the electron pocket is not isotropic, and it leaves little room for a significant hole-like contribution. The contribution of the electron pocket to the normal-state electronic specific heat $C_e$ at $T \to 0$ is[21]: $C_e^{\text{pocket}} / T = (\pi^2 / 2)\,k_B\,(n / T_F) = 5.1 \pm 0.2 \text{ mJ} / \text{mol K}^2$, where $n$ is the carrier density per unit cell (assuming one electron pocket for each of the two $CuO_2$ planes in the unit cell). Recent high-field measurements on a YBCO crystal with $p \approx 0.11$ yield $C_e / T = 4 - 5 \text{ mJ} / \text{mol K}^2$ above 25 T (ref. 28), in agreement with expectation if there are no other sheets in the Fermi surface, or if such sheets have a very low mass $m^*$.

Stripe order is known to be most robust at $p = 1/8$, as indicated for example by the fact that the onset temperature $T_{CO}$ is highest at that doping (see Figure 3b). Hence, the fact that $T_0^S$ in Eu-LSCO and $T_0^H$ in YBCO peak at $p = 1/8$ is further evidence that Fermi-surface reconstruction and stripe order are linked, in both materials. To our knowledge, diffraction studies in YBCO have not yet been carried out in magnetic fields large enough to suppress superconductivity and directly confirm whether or not the non-superconducting ground state in the doping range $0.09 - 0.15$ does indeed have modulations of spin and/or charge. It is conceivable that the stripe-like spin order observed in YBCO at low temperature up to $p \approx 0.07$ in zero field could persist up to higher $p$ when superconductivity is suppressed by a large enough magnetic field[29].

We stress that stripe order extends well above $p = 1/8$, into the highly overdoped region of the phase diagram. The quantum critical point (QCP) where the stripe-ordered phase at $T = 0$ ends on the high-doping side is approximately at $p \approx 0.25$ (refs. 15, 24). Transport measurements[30,31] in Nd-LSCO show clearly that Fermi-surface reconstruction still occurs at $p = 0.20$, but no longer does at $p = 0.24$ (the same is true in Eu-LSCO). The precise location of the QCP is $p^* = 0.235 \pm 0.005$ (ref. 32). The presence of such a QCP has a profound impact on the electronic properties, producing, for example, a linear temperature dependence of the resistivity as $T \rightarrow 0$ (ref. 30), the defining signature of the so-called "strange-metal phase" of cuprates. Significantly, the strength of the anomalous linear-$T$ scattering in hole-doped cuprates was shown to be directly proportional to $T_c$, the strength of superconductivity[2,33], as found in organic superconductors[3].

## Methods

**YBCO samples.** $YBa_2Cu_3O_y$ samples are fully-detwinned crystals grown at the University of British Columbia in non-reactive $BaZrO_3$ crucibles from high-purity starting materials (see ref. 35). The hole concentration (doping) $p$ in each sample is tuned by adjusting the oxygen content $y$. The samples are uncut, unpolished thin platelets, whose transport properties are measured via gold evaporated contacts (of resistance < 1 $\Omega$), in a six-contact geometry. The doping $p$ was determined from a relationship between the superconducting temperature $T_c$ and the $c$-axis lattice constant[36]. The value of $T_c$, defined as the point of zero resistance, is: $T_c = 44.5, 55, 57.3, 61.3$ and 66 K for samples with $y = 6.45, 6.49, 6.51, 6.54$ and 6.67, giving $p = 0.08, 0.09, 0.10, 0.11$ and 0.12, respectively.

**Eu-LSCO samples.** Single crystals of $La_{1.8-x}Eu_{0.2}Sr_xCuO_4$ were grown at the University of Tokyo using a travelling float zone technique. The doping $p$ is taken to be equal to the Sr content $x$, to within $\pm$ 0.005. Electrical contacts (of resistance < 0.1 $\Omega$ at room temperature) were made using silver epoxy diffused into the surface. The value of $T_c$, defined as the point of zero resistance, is: $T_c = 3, 5, 4, 7, 13, 14$ and 9 K for samples with $x = 0.08, 0.10, 0.11, 0.125, 0.16, 0.21$ and 0.24, respectively.

**Measurement of the thermo-electric coefficients.** The thermo-electric transport coefficients were measured by applying a steady heat current through the sample (along the $x$ axis). In YBCO, the current was along the $a$-axis of the orthorhombic crystal structure to avoid any contribution from the CuO chains; in Eu-LSCO, the current flowed within the $CuO_2$ planes. The generated thermal gradient was measured using two uncalibrated Cernox chip thermometers (Lakeshore), referenced to a third, calibrated Cernox. The longitudinal and transverse electric fields were measured using nanovolt preamplifiers and nanovoltmeters. All measurements were performed with the temperature of the experiment stabilized within $\pm$ 10 mK and the magnetic field $B$ swept at a constant rate of 0.4 – 0.9 T / min between positive and negative maximal values, with the heat on. The field was applied normal to the $CuO_2$ planes ($B \parallel z \parallel c$).

Since the Seebeck coefficient $S$ is symmetric with respect to the magnetic field, it is obtained by taking the mean value between positive and negative fields:

$$S = E_x / ( \partial T / \partial x ) = [ \Delta V_x(B) + \Delta V_x(-B) ] / ( 2 \Delta T_x ) \quad ,$$

where $\Delta V_x$ is the difference in the voltage along $x$ measured with and without thermal gradient. This procedure removes any transverse contribution that could appear due to slightly misaligned contacts. The longitudinal voltages and the thermal gradient being measured on the same pair of contacts, no geometric factor is involved.

The Nernst coefficient $N$ is anti-symmetric with respect to the magnetic field, therefore it is obtained by the difference:

$$N = E_y / ( \partial T / \partial x ) = ( L / w ) [ V_y(B) - V_y(-B) ] / ( 2 \Delta T_x ) \quad ,$$

where $L$ and $w$ are the length and width of the sample, respectively along $x$ and $y$, and $V_y$ is the voltage along $y$ measured with the heat current on. This anti-symmetrisation procedure removes any longitudinal thermoelectric contribution from the sample and a constant background from the measurement circuit. The uncertainty on $N$ comes from the uncertainty in determining the sample dimensions, giving typically an error bar of $\pm$ 10 %.

Measurements were performed at the University of Sherbrooke up to 10 or 15 T and at the LNCMI in Grenoble up to 28 T.

## Acknowledgements

We thank C. Kallin, S.A. Kivelson, A.J. Millis, M.R. Norman, P.A. Lee, S. Sachdev, A.-M. Tremblay, and M. Vojta for fruitful discussions. We thank A.B. Antunes and J. Flouquet for their assistance with the experiments at the LNCMI and J. Corbin for his assistance with the experiments at Sherbrooke. J.C. was supported by a Fellowship from the FQRNT and the Swiss SNF. Part of this work was supported by Euromagnet under the EU contract RII3-CT-2004-506239. C.P. and K.B. acknowledge support from the ANR project DELICE. R.L., D.A.B. and W.N.H. acknowledge support from NSERC. L.T. acknowledges support from the Canadian Institute for Advanced Research and funding from NSERC, FQRNT, the Canada Foundation for Innovation, and a Canada Research Chair.

## Author contributions

F.L., J.C., N.D.-L., R.D. and M.R. performed the Nernst and Seebeck measurements in Sherbrooke; F.L., J.C., N.D.-L., E.H., R.D., I.S. and L.M. performed the Nernst and Seebeck measurements at the LNCMI in Grenoble; F.L., J.C., N.D.-L. and L.T. analyzed the data; B.J.R., R.L., D.A.B. and W.N.H. prepared the YBCO samples; S.P., T.T. and H.T. prepared the Eu-LSCO samples; N.D.-L., K.B. and C.P. designed the high-field apparatus for measurements at the LNCMI in Grenoble; F.L. and L.T. wrote the manuscript; L.T. supervised the project.

## Author informations

The authors declare no competing financial interests. Correspondence and requests for materials should be addressed to L.T. (louis.taillefer@physique.usherbrooke.ca).

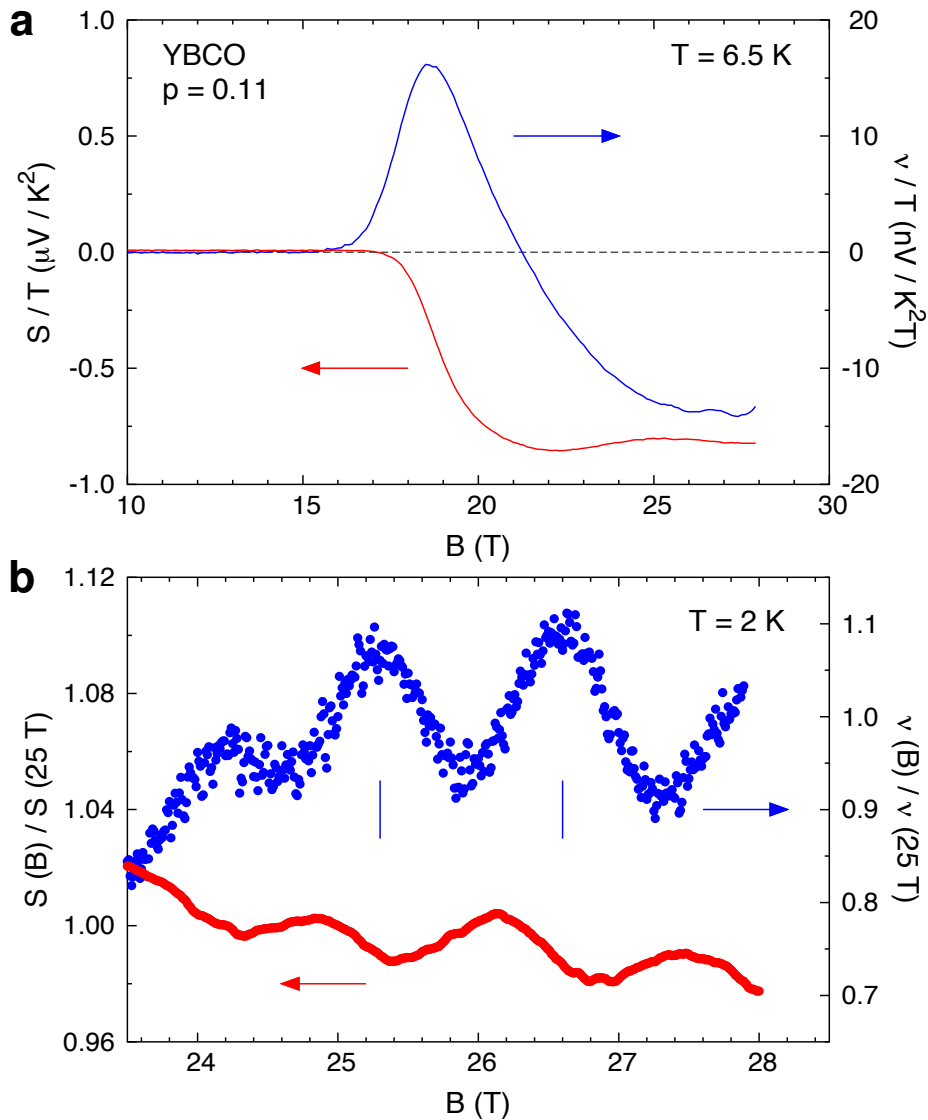

**Figure 1 | Quantum oscillations in the thermo-electric response of YBCO.**

(**a**) Seebeck (*S* ; red curve, left axis) and Nernst (ν ; blue curve, right axis) coefficients of YBCO measured in a single crystal with a doping *p* = 0.11 at a temperature *T* = 6.5 K, potted as *S* / *T* and ν / *T* vs magnetic field *B*. (**b**) Zoom on the high-field range of the same coefficients, measured at *T* = 2 K, normalized to their respective values at *B* = 25 T. Quantum oscillations are clearly seen in both coefficients. The short vertical lines show the spacing of two successive oscillations of frequency *F* = 520 T (in 1/*B*).

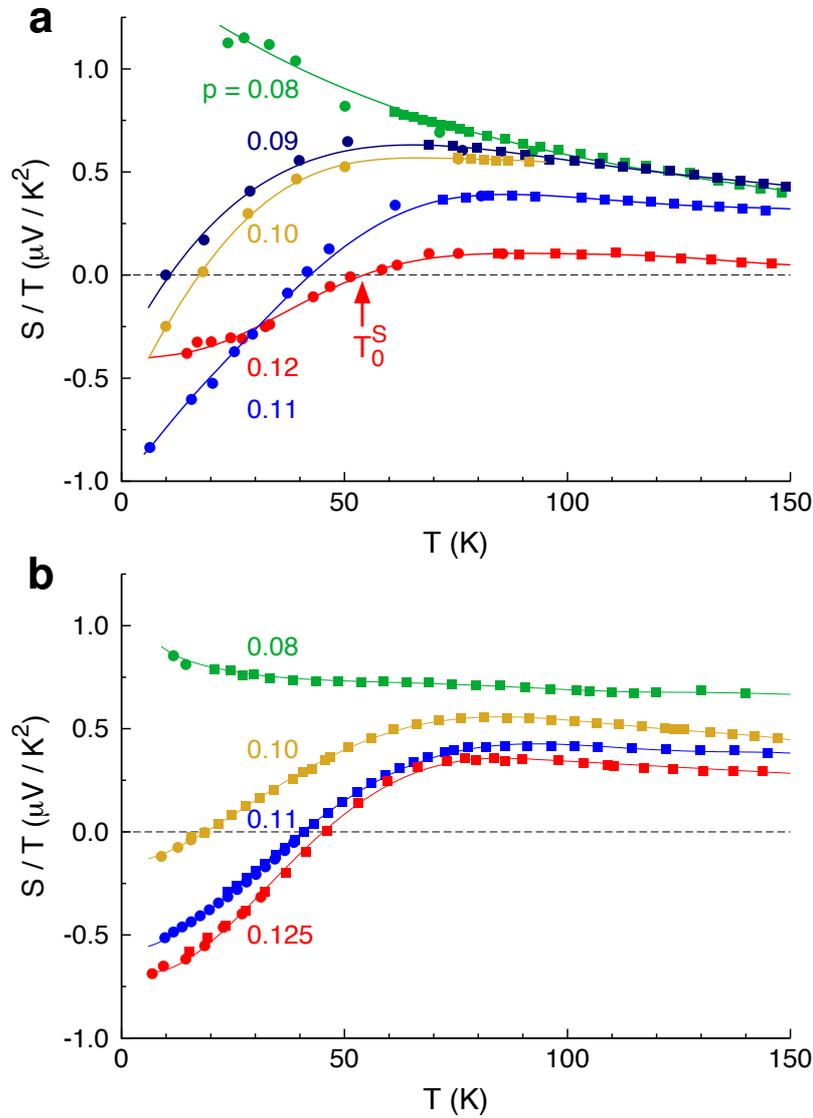

**Figure 2 | Seebeck coefficient of YBCO and Eu-LSCO.**

(**a**) Normal-state Seebeck coefficient of YBCO, plotted as $S / T$ vs $T$, measured in a field $B = 0$ (squares) and $B = 28$ T (circles), at five values of the hole doping $p$ as indicated (data at $p = 0.12$ from ref. 34). $T_0^S$ is the temperature where $S$ changes sign. (**b**) Corresponding data for Eu-LSCO, measured in a field $B = 0$ (squares) and $B = 10$ T (circles), at four dopings as indicated (data at $p = 0.125$ from ref. 34). Our Eu-LSCO data agrees well with previous data[37,38]. All lines are a guide to the eye.

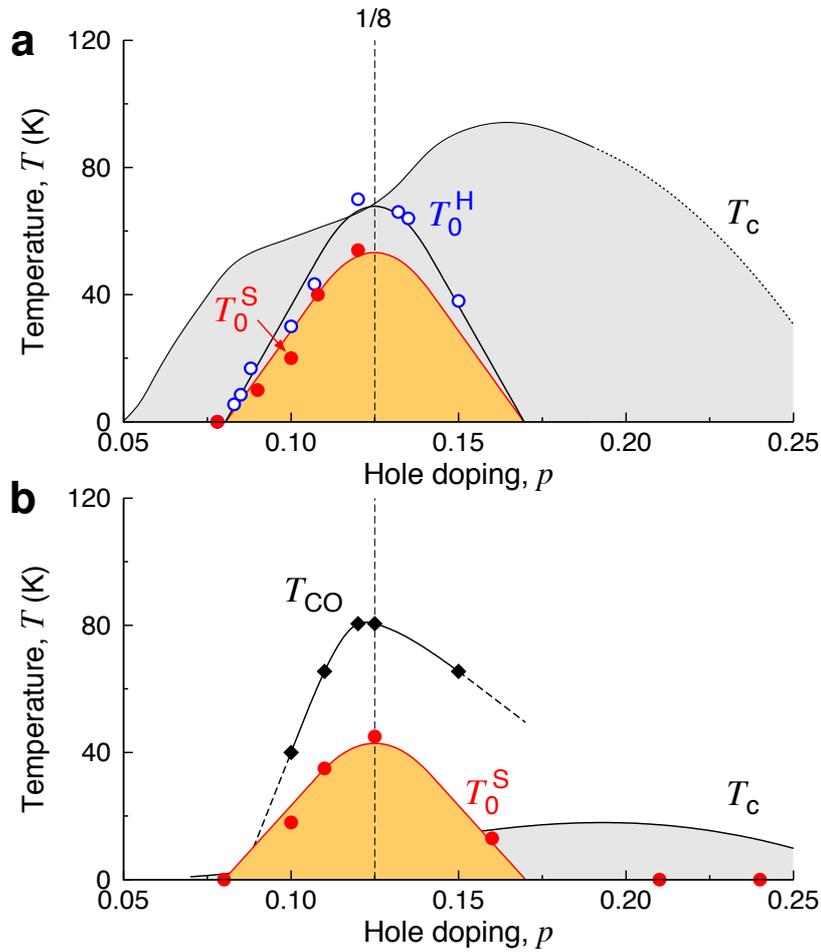

**Figure 3 | Phase diagram of YBCO and Eu-LSCO.**

(**a**) Temperature-doping phase diagram of YBCO showing the zero-field superconducting phase below $T_c$ (grey region) and the region where the normal-state Seebeck coefficient is negative ($S < 0$; yellow dome), delineated by $T_0^S$ (full red circles). Also shown is the line below which the normal-state Hall coefficient is negative ($R_H < 0$), called $T_0^H$ (open blue circles; ref. 20). (**b**) Phase diagram of Eu-LSCO showing $T_c$ (grey region), $T_0^S$ (full red circles) and $T_{CO}$, the onset of charge-stripe order detected by X-ray diffraction (full black diamonds; ref. 15). Lines through $T_0^S$, $T_0^H$ and $T_{CO}$ are a guide to the eye. The black vertical dashed line marks the doping $p = 1/8$.

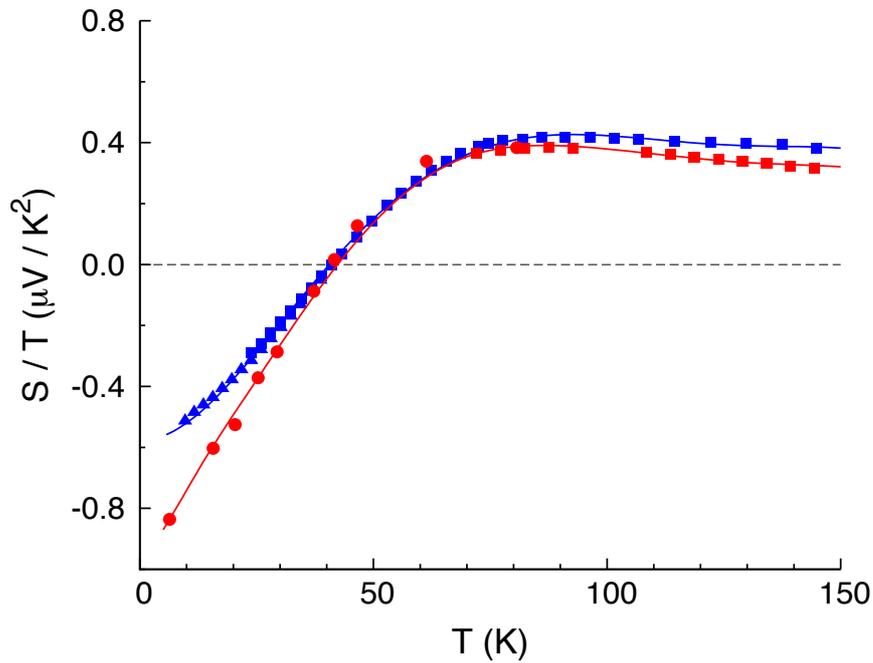

**Figure 4 | Comparing Seebeck data in YBCO and Eu-LSCO.**

Direct comparison of the Seebeck coefficients of YBCO (red symbols) and Eu-LSCO (blue symbols) at the same doping, $p = 0.11$. The applied magnetic field $B = 0$ (squares), 10 T (triangles) or 28 T (circles). The very similar downturn in the normal-state data points to the same Fermi-surface reconstruction.

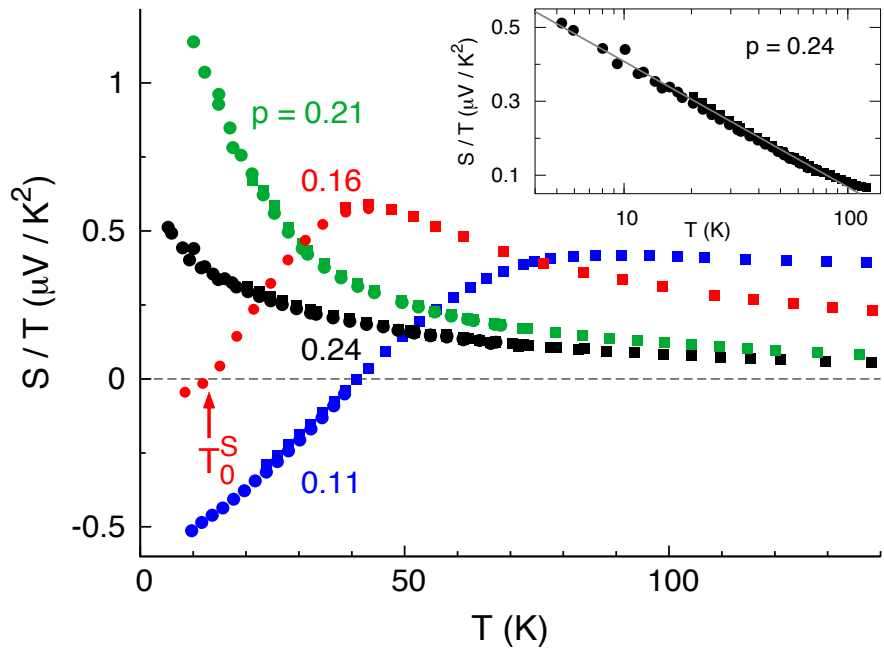

**Figure 5 | Seebeck coefficient of Eu-LSCO at high doping.**

Normal-state Seebeck coefficient $S$ of Eu-LSCO plotted as $S / T$ vs $T$ for values of the hole doping $p$ as indicated, measured in a magnetic field $B = 0$ (squares) and $B = 10$ T (circles). For $p = 0.16$, application of a 10 T field confirms that $S$ becomes negative at low temperature, giving a sign-change temperature $T_0^S \approx 12$ K. For $p = 0.21$ and $p = 0.24$, $S / T$ grows monotonically with decreasing temperature down to the lowest temperature, remaining positive as $T \to 0$, so that $T_0^S = 0$. Inset: $S / T$ vs $\log T$ for $p = 0.24$. The straight line is a good fit to the data from the lowest temperature (5 K) to $\sim 70$ K. This $\log(1/T)$ dependence is a signature of a quantum critical point, also observed in $La_{1.6-x}Nd_{0.4}Sr_xCuO_4$ at $p = 0.24$ (ref. 31).